\documentclass[11pt,aps,pra]{revtex4}
\usepackage[utf8x]{inputenc}
\usepackage{ucs}

\usepackage{amsmath,amssymb}
\usepackage{latexsym}
\usepackage{amsfonts}
\usepackage{dcolumn}
\usepackage{bm}
\usepackage[usenames]{color}
\usepackage{multirow}
\usepackage{graphicx}
\usepackage{hyperref}
\usepackage{subfig}
\usepackage{booktabs}
\usepackage{xcolor}
\usepackage[normalem]{ulem}
\usepackage{float} 
\usepackage{wrapfig} 
\usepackage{upgreek} 
\usepackage{cancel} 
\usepackage{mathdots} 
\usepackage{mathrsfs} 
\usepackage{phaistos}
\usepackage{wasysym}
\usepackage{hieroglf}
\graphicspath{{figures/}} 

\begin{document}

\title{Exact Rényi entropies of $D$-dimensional harmonic systems}


\author{D. Puertas-Centeno, I. V. Toranzo and J. S. Dehesa}
\email[]{dehesa@ugr.es}
\affiliation{Departamento de F\'{\i}sica At\'{o}mica, Molecular y Nuclear, Universidad de Granada, Granada 18071, Spain}
\affiliation{Instituto Carlos I de F\'{\i}sica Te\'orica y Computacional, Universidad de Granada, Granada 18071, Spain}

\begin{abstract}
The determination of the uncertainty measures of multidimensional quantum systems is a relevant issue \textit{per se} and because these measures, which are functionals of the single-particle probability density of the systems, describe numerous fundamental and experimentally accessible physical quantities. However, it is a formidable task (not yet solved, except possibly for the ground and a few lowest-lying energetic states) even for the small bunch of elementary quantum potentials which are used to approximate the mean-field potential of the physical systems. Recently, the dominant term of the Heisenberg and Rényi measures of the multidimensional harmonic system (i.e., a particle moving under the action of a $D$-dimensional quadratic potential, $D > 1$) has been analytically calculated in the high-energy (i.e., Rydberg) and the high-dimensional (i.e., pseudoclassical) limits.
In this work we determine the exact values of the R\'enyi uncertainty measures of the $D$-dimensional harmonic system for all ground and excited quantum states directly in terms of $D$, the potential strength and the hyperquantum numbers.  

\end{abstract}


\maketitle

\section{Introduction}

The R\'enyi entropy of the probability density $\rho(\vec{r}), \vec{r}  =  (x_1 ,  \ldots  , x_D),$ which characterizes the quantum state of a $D$-dimensional physical system is defined \cite{renyi1,renyi2} as
\begin{equation}
\label{eq:renentrop}
R_{q}[\rho] =  \frac{1}{1-q}\log W_{q}[\rho], \quad 0<q<\infty, \,\, q \neq 1,
\end{equation}
where the symbol $W_{q}[\rho]$ denotes the frequency or entropic moment of order $q$ of the density given by
\begin{equation}
\label{eq:entropmom2}
W_{q}[\rho] = \int_{\mathbb{R}^D} [\rho(\vec{r})]^{q}\, d\vec{r}.
\end{equation}
These quantities completely characterize the density $\rho(\vec{r})$ \cite{romera_01,jizba2016} under certain conditions. They quantify numerous facets of the spreading of the quantum probability density $\rho(\vec{r})$, which include the intrinsic randomness (uncertainty) and the geometrical profile of the quantum system. The R\'enyi entropies are closely related to the Tsallis entropies \cite{tsallis} $T_{p}[\rho] = \frac{1}{p-1}(1-W_{p}[\rho]),  0<p<\infty,\, p\neq1$ by $T_{p}[\rho] = \frac{1}{1-p}[e^{(1-p)R_{p}[\rho]}-1]$. Moreover for the special cases $q=0,1,2,$ and $\infty $, the Rényi entropic power, $N_{q}[\rho]=e^{R_{q}[\rho]}$, is equal to $\text{the length of the support}, e^{-\langle \ln \rho \rangle}, \langle \rho \rangle^{-1}, \rho_{max}^{-1}$, respectively.
 Therefore, these $q$-entropies include the Shannon entropy \cite{shannon}, $S[\rho] = \lim_{p\rightarrow 1} R_{p}[\rho] = \lim_{p\rightarrow 1} T_{p}[\rho]$, and the disequilibrium, $\langle\rho\rangle =\exp(\textcolor{red}{-} R_{2}[\rho])$, as two important particular cases; in addition, they 
The use of R\'enyi, Shannon and Tsallis entropies as measures of uncertainty allow a wider quantitative range of applicability than the Heisenberg-like measures which are based on the moments around the origin (so, including the standard or root-square-mean deviation). This permits, for example, a quantitative discussion of quantum uncertainty relations further beyond the conventional Heisenberg-like uncertainty relations \cite{hall,dehesa_sen12,bialynicki2,vignat,zozor2008,guerrero11,puertas2017}. The properties of the Rényi entropies and their applications have been widely analyzed; see e.g. \cite{aczel,leonenko,bialynicki3} and the reviews \cite{dehesa_sen12,jizba,jizba_2004b}. 

In general, the R\'enyi entropies of quantum systems cannot be determined in an exact way, basically because the associated wave equation is generally not solvable in an analytical way. Even when the time-independent Schr\"{o}dinger equation is solvable, what happens for a small set of elementary potentials (zero-range, harmonic, Coulomb) \cite{albeverio,dong2011}, the exact determination of the R\'enyi entropies is a formidable task mainly because they are integral functionals of some special functions of applied mathematics \cite{nikiforov} (e.g., orthogonal polynomials, hypergeometric functions, Bessel functions,...) which control the wavefunctions of the stationary states of the quantum system. These integral functionals have not yet been solved for harmonic (i.e., oscillator-like) systems except for a few lowest-lying states (where the calculation is trivial) and, most recently, for the extreme Rydberg (i.e., highest-lying) \cite{aptekarev2016,dehesa2017,tor2016b} and pseudoclassical (i.e., the highest dimensional) \cite{tor2017b,puertas2017,temme2017} states of harmonic and Coulomb systems by means of sophisticated asymptotical techniques of orthogonal polynomials. This lack is amazing because harmonicity is the most frequent and useful approximation to study the quantum many-body systems, and the other two basic classes of uncertainty measures, the Heisenberg-like measures \cite{ray,drake,hey,assche,andrae,tarasov,zozor,suslov} and the Fisher information \cite{romera2005}, have been already calculated for all stationary states of the multidimensional harmonic system.\\

In this work we determine the exact values of the R\'enyi uncertainty measures of the $D$-dimensional harmonic system (i.e., a particle moving under the action of a quadratic potential) for all ground and excited quantum states directly in terms of $D$, the potential strength and the hyperquantum numbers which characterize the states.
This is a far more difficult problem than the Heisenberg-like and Fisher information cases, both analytically and numerically. The latter is basically because a naive numerical evaluation using quadratures is not convenient due to the increasing number of integrable singularities when the principal hyperquantum number is increasing, which spoils any attempt to achieve reasonable accuracy even for rather small hyperquantum numbers \cite{buyarov}. 

The structure of the manuscript is the following. In section \ref{sec:basics} the wavefunctions and the probability densities of the stationary states of the $D$-dimensional harmonic (oscillator-like) system are briefly described in both position and momentum spaces. In section \ref{sec:renyi} the R\'enyi entropies for all the ground and excited states of this system are determined in an analytical way by use of a recently developed methodology \cite{amc2013}. Finally some conclusions and open problems are given.

\section{The $D$-dimensional harmonic problem}
\label{sec:basics}
In this section we summarize  the quantum-mechanical $D$-dimensional problem corresponding to the harmonic oscillator potential
\begin{equation}
V(r) = \frac{1}{2}k(x_{1}^{2}+\ldots + x_{D}^{2}) = \frac{1}{2}kr^{2},
\end{equation}
and we give the probability densities of the stationary quantum states of the system in both position and momentum spaces. The stationary bound states of the system, which are the physical solutions of the Schr\"{o}dinger equation
\begin{equation}\label{schrodinger}
\left( -\frac{1}{2} \vec{\nabla}^{2}_{D} + V(r)\right) \Psi \left( \vec{r} \right) = E \Psi \left(\vec{r} \right),
\end{equation}
(we use atomic units throughout the paper) where $\vec{\nabla}_{D}$ denotes the $D$-dimensional gradient operator, are well known \cite{gallup,louck60a,yanez1994} to be characterized by the energies
\begin{equation}
\label{HOEL}
E_{N} = \left(N + \frac{D}{2}\right) \omega
\end{equation}
where 
\[
\omega = \sqrt{k}, \quad N = \sum_{i=1}^{D}n_{i} \quad \text{with} \quad n_{\textcolor{red}{i}}=0,1,2,\ldots
\]
The corresponding eigenfunctions can be expressed as
\begin{equation}
\label{HOEF}
\psi_{N}(\vec{r}) = \mathcal{N} e^{-\frac{1}{2}\alpha(x_{1}^{2}+\ldots+x_{D}^{2})}H_{n_{1}}(\sqrt{\alpha}\, x_{1})\cdots H_{n_{D}}(\sqrt{\alpha}\, x_{D}), \quad \alpha = k^{\frac{1}{4}}
\end{equation}	
where $\vec r\in\mathbb R^D$ and $\mathcal{N}$ stands for the normalization constant
\[
\mathcal{N} = \frac{1}{\sqrt{2^{N}n_{1}!n_{2}!\cdots n_{D}! }}\left(\frac{\alpha}{\pi}\right)^{D/4},
\]
and $H_{n}(x)$ denotes the Hermite polynomials of degree $n$ orthogonal with respect the weight function $\omega(x) = e^{-x^{2}}$ in $(-\infty, \infty)$.\\
Then, the associated quantum probability density in position space is given by
\begin{equation}
\label{HOPD}
\rho_{N}(\vec{r}) = |\psi_{N}(\vec{r})|^{2} = \mathcal{N}^{2} e^{-\alpha(x_{1}^{2}+\ldots+x_{D}^{2})}H_{n_{1}}^{2}(\sqrt{\alpha}\, x_{1})\cdots H_{n_{D}}^{2}(\sqrt{\alpha}\, x_{D}),
\end{equation}
and the density function in momentum space is obtained by squaring the Fourier transform of the position wavefunction, obtaining 
\begin{align}
\label{HOMPD}
\gamma_{N}(\vec{p}) & = \mathcal{\tilde{N}}^{2} e^{-\frac{1}{\alpha}(p_{1}^{2}+\ldots+p_{D}^{2})}H_{n_{1}}^{2}\left(\frac{ p_{1}}{\sqrt{\alpha}}\right)\cdots H_{n_{D}}^{2}\left(\frac{ p_{D}}{\sqrt{\alpha}}\right)
= \alpha^{-D}\rho_{N}\left(\frac{\vec{p}}{\alpha}\right)
\end{align}
where $\vec p\in\mathbb R^D$ and the normalization constant is 
\[
\mathcal{\tilde{N}}  = \frac{1}{\sqrt{2^{N}n_{1}!\cdots n_{D}! }}\left(\frac{1}{\pi\alpha}\right)^{D/4}.
\]

\section{R\'enyi entropies of the harmonic system}
\label{sec:renyi}
Let us now determine the R\'enyi entropy of the $D$-dimensional harmonic system according to Eqs. \eqref{eq:renentrop}-\eqref{eq:entropmom2} by
\begin{align}
\label{HORE}
R_{q}[\rho_{N}] &= \frac{1}{1-q}\log \int_{-\infty}^{\infty} dx_{1}\ldots \int_{-\infty}^{\infty} dx_{D} \, [\rho_{N}(\vec{r})]^{q} \nonumber \\
& =  \frac{1}{1-q}\log\left( \mathcal{N}^{2q}\int_{-\infty}^{\infty} e^{-\alpha q x_{1}^{2}}|H_{n_{1}}(\sqrt{\alpha}\, x_{1})|^{2q} \, dx_{1} \cdots \int_{-\infty}^{\infty} e^{-\alpha q x_{D}^{2}}|H_{n_{D}}(\sqrt{\alpha}\, x_{D})|^{2q}\, dx_{D} \right)
\end{align}
where we have used Eq. \eqref{HOPD}. To calculate these $D$ integral functionals of Hermite polynomials we will follow the 2013-dated technique (only valid for $q\in\mathbb{N}$ other than unity) \cite{srivastava,niukkanen,amc2013} to evaluate similar integral functionals of hypergeometric orthogonal polynomials by means of multivariate special functions. To do so, first we express the Hermite polynomials in terms of the Laguerre polynomials (see e.g., \cite{olver}) as
\begin{eqnarray}
\label{HfL}
H_{2n}(x) &=& (-1)^{n} 2^{2n}n!L_{n}^{-\frac{1}{2}}(x^{2}), \nonumber \\
H_{2n+1}(x) &=& (-1)^{n} 2^{2n+1}n!xL_{n}^{\frac{1}{2}}(x^{2}),
\end{eqnarray}
which allows to write
\begin{equation}
\label{HpfL}
H_{n}(\sqrt{\alpha}x) ^{2q} = A_{n,q}(\nu) \alpha^{q\nu}x^{2q\nu} L_{\frac{n-\nu}{2}}^{(\nu-\frac{1}{2})}(\alpha x^{2}) ^{2q},
\end{equation}
with the constant
\[
A_{n,q} (\nu) = 2^{2qn}\left[\Gamma\left(\frac{n-\nu}{2}+1\right) \right]^{2q}
\]
and the paramater $\nu=0(1)$ for even(odd) $n$; that is, $\nu=\frac12\left(1-(-1)^{n}\right).$
\\
Following the same steps as in \cite{amc2013}, after the change of variable $t_{i}=\alpha q x_{i}^{2}$ in \eqref{HORE}, one obtains the following linearization relation for the $(2q)$-th power of the Hermite polynomials 
	\begin{equation}
		\label{linforH2}
		H_{n}\left(\sqrt \alpha x\right)^{2q} = A_{n,q}(\nu)q^{-q\nu}\sum_{j=0}^{\infty}\frac{1}{(-1)^{}2^{2j } j!}c_{j}\left(q\nu,2q,\frac{1}{q},\frac{n-\nu}{2},\nu-\frac{1}{2},-\frac{1}{2} \right) H_{2j}(\sqrt{\alpha q}x),
		\end{equation}
		
with 
\begin{align}
\hspace{-3cm}c_{j}&\left( q\nu,2q,\frac{1}{q},\frac{n-\nu}{2},\nu-\frac{1}{2},-\frac{1}{2} \right)= \nonumber \\
& = \left(\frac{1}{2}\right)_{q\nu} \binom{\frac{n+\nu-1 }{2}}{\frac{n-\nu}{2}}^{2q} F_{A}^{(2q+1)}\left( \begin{array}{cc}
							q\nu+\frac{1}{2} ; \overbrace{\frac{\nu-n }{2}, \ldots, \frac{\nu - n}{2}}^{2q}, -j & \\[-3.5em]
																			&; \underbrace{\frac{1}{q}, \ldots, \frac{1}{q}}_{2q},1\\[-3.5em]
							\underbrace{\nu + \frac{1}{2}, \ldots, \nu+\frac{1}{2}}_{2q},\frac{1}{2} & \\
							\end{array}\right),\nonumber\\
\end{align}
where $(z)_a = \frac{\Gamma(z+a)}{\Gamma(z)}$ is the known Pochhammer's symbol and  $F_{A}^{(2q+1)}(\frac{1}{q}, \ldots, \frac{1}{q},1)$ is the Lauricella function of type A of $2q+1$ variables given by

\begin{align}
\label{HOLF}
F_{A}^{(2q+1)}\left( \begin{array}{cc}
							q\nu+\frac{1}{2} ; \frac{\nu-n }{2}, \ldots, \frac{\nu - n}{2},-j & \\[-3.5em]
																			&; \frac{1}{q}, \ldots, \frac{1}{q},1\\[-3.5em]
							\nu+ \frac{1}{2}, \ldots, \nu+\frac{1}{2},\frac{1}{2} & \\
							\end{array}\right) &=\nonumber\\
							&\hspace{-7cm}=  \sum_{k_{1}, \ldots, k_{2q}, k_{2q+1}=0 }^{\infty} \frac{\left(q\nu+\frac{1}{2}\right)_{k_{1}+\ldots k_{2q}+ k_{2q+1}} (\frac{\nu_-n }{2})_{k_{1}} \cdots (\frac{\nu-n }{2})_{k_{2q}}(-j)_{ k_{2q+1}} }{(\nu+ \frac{1}{2})_{k_{1}} \cdots (\nu + \frac{1}{2})_{k_{2q}}\left(\frac{1}{2}\right)_{ k_{2q+1}} } \frac{\left(\frac{1}{q}\right)^{k_{1}} \cdots \left(\frac{1}{q}\right)^{k_{2q}}}{k_{1}!\cdots k_{2q}!  k_{2q+1}!} ,
\end{align}
 Now, the combination of Eqs. \eqref{HORE} and \eqref{linforH2} together with the orthogonalization condition of the Hermite polynomials $H_{n}(x)$ (with which one realizes that all the summation
terms vanish except the one with $i=0$), allows one to write the exact Rényi entropy of the harmonic system as
\begin{align}
\label{HORE1}
R_{q}[\rho_{N}] &=\frac{1}{1-q}\log \left[\mathcal{N}^{2q} \left(\frac{\pi}{\alpha}\right)^{\frac{D}{2}}q^{-\frac{D}{2}}\prod_{i=1}^{D}q^{-q\nu_{i}}A_{n_{i},q}(\nu_{i})\, c_{0}\left( q\nu_{i},2q,\frac{1}{q},\frac{n_{i}-\nu_{i}}{2},\nu_{i}-\frac{1}{2},-\frac{1}{2} \right)\right]  \nonumber \\
&\hspace{-1cm}= \frac D2 \log\left[\frac{\pi}{\alpha}\right]+\frac{1}{q-1}\log \left[2^{qN} q^\frac D2  \right] +\frac{1}{1-q}\sum_{i=1}^{D}\log\left[\frac{A_{n_{i},q}(\nu_{i})}{q^{q\nu_{i}}\Gamma(n_i+1)^q}\, c_{0}\left( q\nu_{i},2q,\frac{1}{q},\frac{n_{i}-\nu_{i}}{2},\nu_{i}-\frac{1}{2},-\frac{1}{2} \right) \right]
\end{align}
with 
\begin{align}
c_{0}&\left( q\nu,2q,\frac{1}{q},\frac{n-\nu}{2},\nu-\frac{1}{2},-\frac{1}{2} \right) = \left(\frac{1}{2}\right)_{q\nu} \binom{\frac{n+\nu-1 }{2}}{\frac{n-\nu}{2}}^{2q}\, \mathfrak F_q(n),\nonumber\\
\end{align}
where the symbol $\mathfrak F_q(n)$ denotes the following Lauricella function  of $2q$ variables
\begin{align}
\label{HOLF2}
\mathfrak F_q(n)&\equiv 
F_{A}^{(2q+1)}\left( \begin{array}{cc}
							q\nu+\frac{1}{2} ; \frac{\nu-n }{2}, \ldots, \frac{\nu - n}{2},0 & \\[-3.5em]
																			&; \frac{1}{q}, \ldots, \frac{1}{q},1\\[-3.5em]
							\nu+ \frac{1}{2}, \ldots, \nu+\frac{1}{2},\frac{1}{2} & \\
							\end{array}\right) =
F_{A}^{(2q)}\left( \begin{array}{cc}
							q\nu+\frac{1}{2} ; \frac{\nu-n }{2}, \ldots, \frac{\nu - n}{2} & \\[-3.5em]
																			&; \frac{1}{q}, \ldots, \frac{1}{q}\\[-3.5em]
							\nu + \frac{1}{2}, \ldots, \nu +\frac{1}{2} & \\
							\end{array}\right) &\nonumber\\
							&=  \sum_{j_{1}, \ldots, j_{2q}=0 }^{\infty} \frac{\left(q\nu+\frac{1}{2}\right)_{j_{1}+\ldots j_{2q}} (\frac{\nu-n }{2})_{j_{1}} \cdots (\frac{\nu-n }{2})_{j_{2q}} }{(\nu + \frac{1}{2})_{j_{1}} \cdots (\nu+ \frac{1}{2})_{j_{2q}} } \frac{\left(\frac{1}{q}\right)^{j_{1}} \cdots \left(\frac{1}{q}\right)^{j_{2q}}}{j_{1}!\cdots j_{2q}! }&
							\nonumber\\
							&=  \sum_{j_{1}, \ldots, j_{2q}=0 }^{\frac{n-\nu}2} \frac{\left(q\nu+\frac{1}{2}\right)_{j_{1}+\ldots j_{2q}} (\frac{\nu-n }{2})_{j_{1}} \cdots (\frac{\nu-n }{2})_{j_{2q}} }{(\nu + \frac{1}{2})_{j_{1}} \cdots (\nu + \frac{1}{2})_{j_{2q}} } \frac{\left(\frac{1}{q}\right)^{j_{1}} \cdots \left(\frac{1}{q}\right)^{j_{2q}}}{j_{1}!\cdots j_{2q}! }.
\end{align}
Note that, as $\frac{\nu-n}{2}$ is always a negative integer number, the Lauricella function simplifies to a finite sum. In the following, for convenience, we use the notation $N_O=\sum_{i=1}^D\nu_i$, which is the amount of odd numbers $n_i$ and, thus, $N_E=D-N_O$ gives the number of the even ones. Then simple algebraic manipulations allow us to rewrite Eq. \eqref{HORE1} as
\begin{eqnarray}
\label{HORE3}\nonumber
R_{q}[\rho_{N}] 
&=&    -\frac D2 \log\left[\alpha\right]+\mathcal K_q\,D+\overline{\mathcal K}_q\,N_O+\frac {q}{q-1}\sum_{i=1}^D(-1)^{n_i}\log\left[\left(\frac{n_i+1}{2}\right)_{\frac12}  \right]+\frac{1}{1-q}\sum_{i=1}^D\log\left[\mathfrak F_q(n_i)\right], \nonumber\\
\end{eqnarray}
where $\mathcal K_q=\frac{\log[\pi^{q-\frac12}\,q^\frac12]}{q-1}$ and $\overline{\mathcal K}_q=\frac{1 }{1-q}\log \left[\frac{4^{q}\,\Gamma\left(\frac12+q\right)}{\pi^{\frac{1}{2}}\,q^{q}}\right]$.
This expression allows for the analytical determination of the Rényi entropies (with positive integer values of $q$) for any arbitrary state of the multidimensional harmonic systems.

Finally, for the ground state (i.e., $n_i=0,\,i=1,\cdots, D$; so, $N=0$) the general Eq. \eqref{HORE3} boils down to , 
	 	 \begin{equation}
	 	 R_q[\rho_N]=\frac D2\log\left[\frac {\pi\, q^{\frac1{q-1}}}{\alpha}\right].
	 	 \end{equation}
In fact, this ground state R\'enyi entropy holds for any $q>0$ as one can directly derive from Eq. \eqref{HORE}. Taking into account that the momentum density is a re-scaled form of the position density, we have the following expression for the associated momentum R\'enyi entropy,
\begin{eqnarray}
\label{Remsp}
R_{\tilde q}[\gamma_N] &=&  \frac D2 \log\left[\alpha\right]+\mathcal K_{\tilde q}\,D+\overline{\mathcal K}_{\tilde q}\,N_O+\frac {\tilde q}{\tilde q-1}\sum_{i=1}^D(-1)^{n_i}\log\left[\left(\frac{n_i+1}{2}\right)_{\frac12}  \right]+\frac{1}{1-\tilde q}\sum_{i=1}^D\log\left[\mathfrak F_{\tilde q}(n_i)\right],\nonumber\\
\end{eqnarray}
($\tilde q\in \mathbb{N}$). Although Eqs. \eqref{HORE3} and \eqref{Remsp} rigorously hold for $q\not=1$ and $q\in\mathbb N$ only, it seems reasonable to conjecture its general validity for any $q>0, \,q\not=1$ provided the formal existence of a generalized function $\mathfrak F_q(n)$. If so, we obtain the general expression for the position-momentum uncertainty Rényi entropic sum as
\begin{eqnarray}\nonumber
R_{q}[\rho_N]+R_{\tilde q}[\gamma_N] &=&  (\mathcal K_{q}+\mathcal K_{\tilde q})\,D+(\overline{\mathcal K}_{q}+\overline{\mathcal K}_{\tilde q})\,N_O+\left(\frac {q}{ q-1}+\frac {\tilde q}{\tilde q-1}\right)\sum_{i=1}^D(-1)^{n_i}\log\left[\left(\frac{n_i+1}{2}\right)_{\frac12}  \right]
\\
&+&\frac{1}{1- q}\sum_{i=1}^D\log\left[\mathfrak F_{q}(n_i)\right]+\frac{1}{1-\tilde q}\sum_{i=1}^D\log\left[\mathfrak F_{\tilde q}(n_i)\right]
\end{eqnarray}
which verifies the R\'enyi-entropy-based uncertainty relation of Zozor-Portesi-Vignat \cite{zozor2008} when $\frac1q+\frac1{\tilde q}\ge2$ for arbitrary quantum systems.
In the conjugated case $\tilde q=q^*$ such that $\frac1q+\frac1{q^*}=2$, one obtains
\begin{eqnarray}\nonumber
R_{q}[\rho_N]+R_{q^*}[\gamma_N] &=&  D\log\left(\pi q^{\frac1{2q-2}}{q^*}^{\frac{1}{2q^*-2}}\right)+(\overline{\mathcal K}_{q}+\overline{\mathcal K}_{ q^*})\,N_O\\
&+&\frac{1}{1- q}\sum_{i=1}^D\log\left[\mathfrak F_{q}(n_i)\right]+\frac{1}{1- q^*}\sum_{i=1}^D\log\left[\mathfrak F_{ q^*}(n_i)\right].
\end{eqnarray}
Let us finally remark that the first term corresponds to the sharp bound for the general Rényi entropy uncertainty relation with conjugated parameters
\begin{equation}\nonumber
R_{q}[\rho_N]+R_{q^*}[\gamma_N] \ge D\log\left(\pi q^{\frac1{2q-2}}{q^*}^{\frac{1}{2q^*-2}}\right)
\end{equation}
of Bialynicki-Birula \cite{bialynicki2} and Zozor-Vignat \cite{vignat}.

\section{Conclusions}

In this work we have explicitly calculated the R\'enyi entropies, $R_q [\rho_N]$ ($q\in \mathbb{N}$), for all the 
quantum-mechanically allowed harmonic states in terms of the Rényi index $q$, the spatial dimension $D$, the oscillator strength $\alpha$, as well as
the hyperquantum numbers, $\{n_{i}\}_{i=1}^{D}$, which characterize the corresponding state's wavefunction. To do that we have used the harmonic wavefunctions in Cartesian coordinates, which can be expressed in terms of a product of $D$ Hermite polynomials and exponentials. So, the R\'enyi entropies of the quantum states boil down to $D$ entropy-like functionals of Hermite polynomials. Then we have determined these integral functionals by taking into account the close connection between the Hermite and Laguerre polynomials and the Srivastava-Niukkanen linearization method for powers of Laguerre polynomials. The final analytical expression of the Rényi entropies with positive integer index $q$ in both position and momentum spaces is given in a compact way by use of a Lauricella function of type A. It remains as an open problem, the extension of this result to R\'enyi entropies for any real value of the parameter $q$. The latter requires a completely different approach, still unknown to the best of our knowledge. 

\section*{Acknowledgments}
This work has been partially supported by the Project FQM-207 of the Junta de Andaluc\'ia and the MINECO-FEDER grants FIS2014-54497P and FIS2014-59311P. I. V. Toranzo acknowledges the support of ME under the program FPU.
\\
Author contribution statement:
all authors have contributed equally to the paper.

 \end{document}